\documentclass[letterpaper,10pt,twocolumn,final,conference,oneside]{IEEEtran}
\usepackage{array}
\usepackage{epsfig}
\usepackage{graphics}
\usepackage{graphicx}
\usepackage[latin1]{inputenc}
\usepackage{amsmath}
\usepackage{amssymb}
\usepackage{booktabs}
\usepackage{subfigure}
\usepackage{bm}
\usepackage{cite}
\usepackage{cases}
\usepackage{color,soul}
\usepackage{multirow}
\usepackage{balance}
\usepackage{acronym}
\usepackage[hidelinks]{hyperref}

\usepackage{tikz}

\acrodef{wgsr}[WGSR]{water gas shift reactor}
\acrodef{bess}[BESS]{battery energy storage system}
\acrodef{pms}[PMS]{power management system}
\acrodef{mgcc}[MGCC]{microgrid central controller}
\acrodef{soc}[SOC]{state of charge}
\acrodef{dg}[DG]{diesel generator}
\acrodef{vsi}[VSI]{voltage source inverter}
\acrodef{csi}[CSI]{current source inverter}
\acrodef{avr}[AVR]{automatic voltage regulator}
\acrodef{ems}[EMS]{energy management system}
\acrodef{der}[DER]{distributed energy resource}
\acrodef{mcr}[MCR]{maximum continuous rating}
\acrodef{pv}[PV]{photovoltaic}
\acrodef{milp}[MILP]{mixed-integer linear programming}

\IEEEoverridecommandlockouts

\newcommand\copyrighttext{%
  \footnotesize
  \centering\copyright~2022 IEEE. Personal use of this material is permitted. Permission from IEEE must be obtained for all other uses, in any current or future media, including reprinting/republishing this material for advertising or promotional purposes, creating new collective works, for resale or redistribution to servers or lists, or reuse of any copyrighted component of this work in other works.\\
  IEEE PES General Meeting 2022. 
  DOI:\href{https://doi.org/10.1109/PESGM48719.2022.9916899}{10.1109/PESGM48719.2022.9916899}}
\newcommand\copyrightnotice{%
\begin{tikzpicture}[remember picture,overlay]
\node[anchor=south,yshift=10pt] at (current page.south) {\setlength{\fboxrule}{0pt}\fbox{\parbox{\dimexpr\textwidth-\fboxsep-\fboxrule\relax}{\copyrighttext}}};
\end{tikzpicture}%
}

\usepackage{siunitx}
\DeclareSIUnit{\atm}{atm}
\DeclareSIUnit{\kWh}{kWh}
\DeclareSIUnit{\Ah}{Ah}
\DeclareSIUnit{\MWh}{MWh}
\DeclareSIUnit{\MWp}{MWp}
\DeclareSIUnit{\MVA}{MVA}

\begin{document}

\title{An Efficiency-Based Power Management Strategy for an Isolated Microgrid Project}

\author{%
 \IEEEauthorblockN{Francesco Conte, Fabio D'Agostino,\\ Gabriele Mosaico, Federico Silvestro}
 \IEEEauthorblockA{University of Genoa\\
    DITEN\\
    Via all'Opera Pia 11 A\\
    I-16145, Genova, Italy\\
    fabio.dagostino@unige.it}
 \and
  \IEEEauthorblockN{Samuele Grillo}
  \IEEEauthorblockA{\\Politecnico di Milano\\
    DEIB\\
    p.zza Leonardo da Vinci, 32\\
    I-20133, Milano, Italy\\
    samuele.grillo@polimi.it}}

\IEEEaftertitletext{\copyrightnotice\vspace{0.2\baselineskip}}
\maketitle

\begin{abstract}
The microgrids design for remote locations represents one of the most important and critical applications of the microgrid concept. It requires the correct sizing and the proper utilization of the different sources to guarantee the economical feasibility and the reliability of the supply. This study illustrates an efficiency-based power management strategy,  designed for an undergoing microgrid project, where the sizing process of the resources (diesel generators, battery energy storage system, and PV plant) is obtained using a mixed-integer optimization algorithm.
The proposed power management strategy guarantees the efficient exploitation of the power sources, which is one of the key elements of the optimal sizing process, being naturally included in the definition of the energy cost functions.
The effectiveness of the power control strategy is validated by means of quasi-dynamic simulations on the complete microgrid model, where sources are defined by the optimal problem solution, while the cabling (size and length) and the main switchboards location reflect the expected system layout.
Results obtained from the simulation of the microgrid electrical system include losses, and allow to verify and to highlight the desired quantities, such as the quality of supply at each busbar (voltage magnitude), and the state of charge of the energy storage system.
\end{abstract}


\section{Introduction}\label{sec:Introduction}
In the last years the energy sector has undergone a tremendous change. A wave of growing demography in emerging economies, tightening regulatory pressures in developed economies, advancements in technology and new business models have all supported a long-lasting transformation.

The spread of \acp{der}, especially photovoltaic and wind, has changed radically the electrical system. Before the advent of distributed generation, the energy was produced in a few large thermal, hydro or nuclear plants and it was transmitted and distributed to the load in a radial and unidirectional power flow. After the coming of DG, the electrical system must face with lots of little conventional and renewable distributed generators.
Modern power systems are called to manage the bidirectional power flows and the spread of renewable generation, which has the main characteristic of being stochastic and not dispatchable~\cite{Grillo:2012}. This is a significant challenge when trying to guarantee the balance between load demand and power generation. This problem is much more important when the size of the grids shrinks as distribution systems move from a centralized paradigm towards a local, decentralized, and potentially islanded operation \cite{Phan:2019}.
Four major trends are at the foundation of this recent evolution and will continue to play in the decades ahead: i) diffusion of \acp{der}; ii) a strong commitment to increase energy efficiency; iii) digitalisation of the grid; and iv) energy storage development.

All these factors made the big centralized electrical system move to more decentralized and local ones; in this scenario microgrids are one of the most interesting challenges \cite{Lopes:2013}.
A microgrid is an energy supply network built around local power and (usually also) heat generation facilities \cite{Stevanato:2020}, and it is designed to operate autonomously or in parallel with a national grid  \cite{Cagnano:2019}. In order to foster economy growth, it is vital to develop local microgrids capable of generating and distributing electricity in many contexts, such as islands, and remote areas. The same solution provides the additional opportunity to benefit from clean, renewable energy sources \cite{Kiptoo:2019,Worighi:2019}.
A microgrid is typically made up of: i) renewable energy sources (solar, wind or biomass); ii) fossil fuel energy sources (conventional sources) to ensure grid stability; iii)  energy storage solutions (batteries, hydrogen storage, mechanical storage, etc.); and iv)  a low-voltage distribution grid~\cite{dago:2016}.

Microgrids have become competitive as a result of technological progress and decreasing prices of certain key components, including photovoltaic panels, batteries and control systems~\cite{Hatziargyriou:2020}, together with the implementations of energy management strategies \cite{Vera:2019,Ahmad:2020}.
In order to properly manage the installed units and the power flows during operation, the \ac{mgcc} has a fundamental role in both islanded and grid connected operating conditions \cite{Olivares:2014,Abdelaziz:2016}. It is also of paramount importance to properly design the microgrid, setting the right sizes of the various equipment \cite{Shirzadi:2020}, especially the \acp{bess} \cite{Carpinelli:2018}.

The aim of this paper is twofold. From the one hand it describes a mixed integer linear programming-based methodology to optimally design the microgrid. On the other hand, it describes a control methodology of the microgrid in which the \acp{bess} are the grid-forming units. The rationale of this choice is the fact that the main scope of the \ac{mgcc} is to reduce the costs from fossil-fuel-based generation and to maximize the exploitation of the energy from renewable sources. The contribution of this paper is that the decision on the power management is subject to an efficiency-based logic and the round-trip efficiency of the \acp{bess} is used as a threshold to determine whether or not the \acp{dg} should be activated or switched off.

The paper is organized as follows. Section~II describes the optimization algorithm along with the obtained configuration. Section~III is devoted to the description of the \ac{pms} and the results of a 1-year quasi-dynamic simulation of the microgrid are analyzed in Section~IV. Finally, conclusions are drawn in Section~V.

\section{Microgrid Optimal Sizing}

The optimal sizing is performed using a \ac{milp} formulation, presented in this Section. It is implemented in Matlab-AMPL environment, using CPLEX solver. The complete formulation is omitted in order to focus on the most innovative aspects of this work, but the reference model can be found in~\cite{Microgrid_ABB} (with due adjustments to unknown sizes of devices).

\subsection{Methodology}

The sizing is performed on a single busbar microgrid. The main constraint is the classical electrical balance equation
\begin{equation} \label{eq:balance}
\sum_L P_{\rm L}(t) = \sum_g P_{\rm g}(t) + \sum_b P_{\rm b}(t) + P_{\rm PV}(t) - P_{\rm curt}(t),
\end{equation}
where $P_{\rm L}$ is the load power profile, $P_{\rm g}$ and $P_{\rm PV}$ are the power profiles delivered by \ac{dg} units and by the \ac{pv} plant respectively, and  $P_{\rm b}$ is the power profile exchanged by the \acp{bess}. A curtailment  $P_{\rm curt}$ of the \ac{pv} plant is allowed, though \textit{a priori} could be not convenient, in order to avoid the oversizing of \acp{bess} to guarantee the balance even in off-grid operation.

The objective function is composed by the capital expenditures for all the components to be installed, the operating expenses of diesel (fuel consumption and maintenance). Capital expenditure of photovoltaic plant and \acp{bess} is surcharged in order to include maintenance costs, assumed to be  proportional to rated capacity. A pay-back time of 8 years is imposed for the investments above.
Thus, the object function can be written as
\begin{equation} \label{eq:obj}
C_{\rm g} P_{\rm g}^{\rm n} + C_{\rm Pb} P_{\rm b}^{\rm n} + C_{\rm Eb} E_{\rm b}^{\rm n} + C_{\rm PV} P_{\rm PV}^{\rm n} + \sum_{t=1}^{T} f \left( P_{\rm g}(t) \right),
\end{equation}
where the size of each device ($P_{\rm g}^{\rm n}$ of \acp{dg}, $P_{\rm b}^{\rm n}$ and $E_{\rm b}^{\rm n}$ for energy capacity and power rating of storage devices and $P_{\rm PV}^{\rm n}$ for \ac{pv}) is weighted by the corresponding costs ($C_{\rm g}$, $C_{\rm Pb}$ and $C_{\rm Eb}$, $C_{\rm PV}$), plus fuel consumption cost $f$.

All the controllable devices are then modelled with proper constraints:
\subsubsection{Diesel generators}
They must respect minimum and maximum active power limits, proportional to the (unknown) number of devices installed. Three discrete sizes of DGs (\SI{1.9}{\MVA}, \SI{2.5}{\MVA}, and \SI{3}{\MVA}, i.e., \SI{1.5}{\mega\watt}, \SI{2}{\mega\watt}, and \SI{2.4}{\mega\watt}) can be chosen by the algorithm.
Fuel consumption curve is a linear function of active power setpoint~\cite{DeSerio2017,HomerDiesel}, not just proportional to it, but with a y-intercept that leads to an hyperbolic efficiency curve. A binary variable to represent the on/off status of the device is necessary.
In order to reduce the computational burden and considering the fact that the aim of this phase is the planning, start up cost, ramp limits, and minimum up and down times are neglected.
\subsubsection{Storage systems}
They are more complex: in addition to power limitations, they require \ac{soc} computation (thus, an inter-temporal constraint), which is bound to stay between 30\% and 90\% of the (unknown) rated capacity.
Two different values of C-rate (the ratio of rated active power to the nominal energy, i.e., the maximum constant power that the battery can exchange in an hour) are available, 1C and 0.5C, and the algorithm chooses the most appropriate, in terms of objective function minimization.
Charge and discharge efficiencies are modelled according to~\cite{Conte_BESS} (respectively 0.87 and 0.86), using binary variables to identify charge and discharge states.
\subsubsection{Photovoltaic plant}
Actually this component is not controllable by the system: its production, $P_{\rm PV}(t)$, is just proportional to input profiles (described in next subsection). Curtailment power $P_{\rm curt}(t)$ is a slack variable, positive and bound to be always smaller than $P_{\rm PV}(t)$.

\subsection{Inputs} \label{section:inputs}
The main input of the problem is the electrical demand to be supplied. Typical load profiles have been developed in order to simulate a 10-year horizon, combining typical daily and annual profiles for a resort, scaled on a \SI{6}{\mega\watt}-rated load, with random fluctuations around average trends.

Real measurements of local solar irradiance (in a time span 2005--2014) are used and converted into the expected \ac{pv} power production according to~\cite{Mosaico_PV}:
\begin{equation}
P_{\rm PV}(t)= \frac{I_{\rm g}(t) P_{\rm PV}^{\rm n} \eta_{\rm pan}(t) \eta_{\rm inv} \omega_{\rm deg}(t)}{I_{\rm STD}},
\end{equation}
where $I_{\rm g}$ is the global horizontal irradiance, $P_{\rm PV}^{\rm n}$ is the total rated power (unknown here, thus the production is scaled proportionally to this value), $I_{\rm STD}$ is the (constant) irradiance at standard test conditions, $\eta_{\rm pan}(t)$ and $\eta_{\rm inv}$ are the efficiencies of the panels and inverter, and $\omega_{\rm deg}(t)$ is a degradation coefficient, well detailed in ~\cite{Mosaico_PV}.

\begin{figure}[!ht]
    \centering
    \includegraphics[width=0.7\columnwidth]{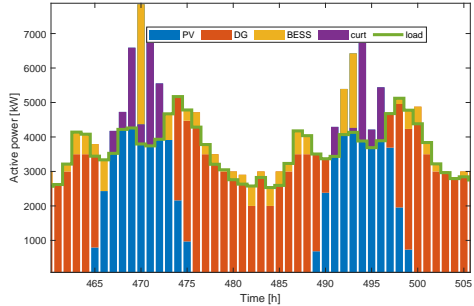}
    \caption{Detail of the microgrid power balance performed by the sizing optimization code.}
    \label{fig:sizing}
\end{figure}

\subsection{Results}
Optimal sizes obtained with the proposed formulation are:
\begin{itemize}
    \item \ac{pv} = \SI{10}{\MWp}
    \item \ac{bess} = \SI{5}{\mega\watt}/\SI{5}{\MWh}
    \item Two \SI{1.9}{\MVA}-rated and one \SI{2.5}{\MVA}-rated diesel generators.
\end{itemize}
In Figure~\ref{fig:sizing}, a focus on two daily power profiles of the microgrid components is offered, highlighting the correct match of demand (green line) with the mix of production, storage and curtailment. BESS power is to be considered an absorption (charging) when it is above the load line, or a power injection (discharging) when below that line.

The obtained configuration is compared with that obtained by HOMER (Hybrid Optimization of Multiple Energy Resources) software using the same economical and technological assumptions. The resulting configuration is:
\begin{itemize}
    \item \ac{pv} = \SI{10}{\MWp}
    \item \ac{bess} = \SI{5}{\mega\watt}/\SI{24}{\MWh}
    \item Two \SI{1.9}{\MVA}-rated, one \SI{2.5}{\MVA}-rated and one \SI{1.25}{\MVA}-rated diesel generators.
\end{itemize}
Thus, it can be stated that a significant reduction is reached by the proposed technique.
Even if it suggests less and smaller devices, it is verified that all the constraints are met throughout the whole simulated horizon and the overall system is more efficient.

\subsection{Grid layout}

\begin{figure}[!ht]
    \centering
    \includegraphics[width=0.9\columnwidth]{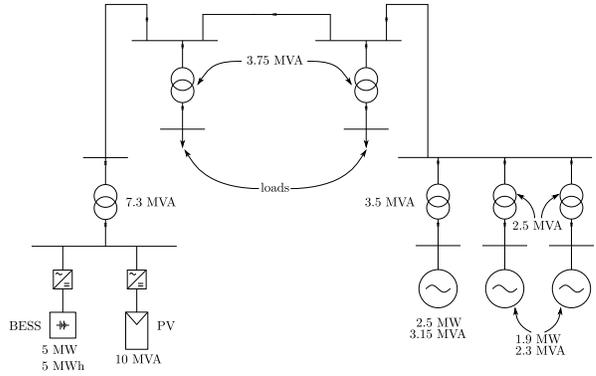}
    \caption{One-line diagram of the microgrid resulting from the optimization algorithm and used for tests.}
    \label{fig:topology}
\end{figure}

The grid topology is depicted in Figure~\ref{fig:topology}. It has been defined with the aim to respect geographical constrains of the prospected location (an island), while ensuring reliable supply to loads, by decoupling the two sets of generators (PV, and \acp{dg}). Transformers, lines and electrical machines have been modeled and characterized using parameters from literature and technical data sheets, in order to reproduce a valid simulation model. The microgrid has been implemented in DIgSILENT PowerFactory.

\section{Power Management Strategy}
The power management strategy has been designed to satisfy the operational limits of components, while maximizing the energy production efficiency. The consolidated microgrid control system architecture refers to a well known hierarchical structure~\cite{Guerrero2011,Jackson:2020}, where local controllers are coordinated by the \ac{mgcc}. The \ac{mgcc} implements the \ac{pms} that has the main task to ensure the power availability at each time, by managing generators in order to avoid black-out occurrences and to maintain the quality of the supply at load busbars. The \ac{pms} ensures adequate primary reserve, switches on and off generators and realizes the power sharing among controllable generating units.

\subsection{Local controllers}
Each power source is equipped with its own local controller. \ac{bess} represents the grid forming unit, and it regulates system frequency and voltage magnitude at the generation busbar. It can be modeled with a \ac{vsi}~\cite{Rocabert:2012} and in quasi-dynamic simulations is the \textit{slack} node.

\acp{dg} adopted in this study are controlled as constant power factor sources. Their representation in quasi-dynamic simulations corresponds to a \textit{PQ} node.

The PV plant is a grid feeding unit, well described by a \ac{csi} model, whose power injection is driven by the solar irradiance. The PV plant can be only partially controlled and offer very limited regulation capabilities to the system (i.e., operation at constant power factor and power curtailment functionality). Its representation for quasi-dynamic simulations corresponds to a \textit{PQ} node.

\subsection{Power Management System}
The essential role of a \ac{pms} for storage-based isolated microgrids is to maintain the \ac{soc} of the BESS within the operational limits. While the optimization procedure has the perfect knowledge of load and PV production profiles over the entire time horizon, the \ac{pms} works without any prior information concerning future power absorption and production. Energy management functionalities based on forecasting tools are not included in this work, being this power management algorithm designed to be implemented on-board commercial PLC or intelligent breakers with limited computational capabilities.

\begin{figure}[ht]
\centering
\includegraphics[width=\columnwidth]{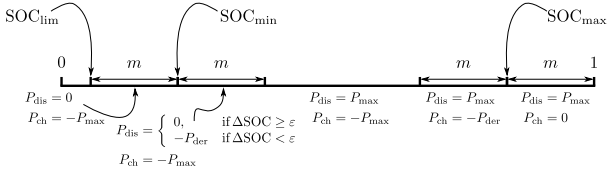}
\caption{\ac{pms} control variable definition.}
\label{fig:SoC}
\end{figure}

The first task of the \ac{pms} is to define, for each control interval, the available charging  ($P_{\rm ch}$) and discharging ($P_{\rm dis}$) powers of the \ac{bess}, according to the current \ac{soc} value, as shown in Figure~\ref{fig:SoC}.

In Figure~\ref{fig:SoC} ${\rm SOC}_{\rm min}$ and ${\rm SOC}_{\rm max}$ are the minimum and maximum \ac{soc} values, defined as
\setlength{\arraycolsep}{0.0em}
\begin{eqnarray}
{\rm SOC}_{\rm min} &=& {\rm SOC}_{\rm lim} + m\\\label{eq:soc1}
{\rm SOC}_{\rm max} &=& 1 - m\label{eq:soc2}
\end{eqnarray}\setlength{\arraycolsep}{5pt}
where the margin value $m$ allows the battery to maintain a residual amount of regulating energy ensuring the primary reserve. The limit value ${\rm SOC}_{\rm lim}$ is equal to 0.2, and represents the usual lower operational limit of Li-ion-based storage systems. Once the margin is decided by the microgrid operator, the derating power ($P_{\rm der}$) is automatically defined as
\begin{equation}
    P_{\rm der} = \frac{m E_{\rm BESS}^{\rm n}}{\eta_{\rm BESS} T_{\rm ctrl}/60}
   \label{eq:pdor}
\end{equation}
where:
\begin{itemize}
    \item $E_{\rm BESS}^{\rm n}$ is the nominal energy of the BESS;
    \item $T_{\rm ctrl}$ is the time control interval of the \ac{pms} (in \si{\min});
    \item $\eta_{\rm BESS}$  is the global charging/discharging efficiency.
\end{itemize}

The derating value corresponds to the active power production (or absorption) that allows to respect the operational limits of the battery, even if the actual state of charge is close to either upper or lower bounds.

It is important to notice that when the \ac{soc} is within the thresholds ${\rm SOC}_{\rm min}$ and ${\rm SOC}_{\rm min} + m$, the discharging power value depends also on the variation of the \ac{soc} itself ($\Delta {\rm SOC}$). This peculiarity allows the \ac{pms} to avoid the interruption of a charging process while ensuring the full discharging capability of the battery when it provides power to the system.

The proposed power management algorithm is based on a state-space representation of the microgrid operating conditions. The \ac{pms} does not control the active power reference of the \ac{bess}, which is the grid-forming unit, but controls the diesel generators set-points. The key control variable of the power control strategy is given by
\begin{equation}
    P_{\rm ctrl}= P_{\rm L} - P_{\rm PV} - P_{\rm dis}
    \label{eq:kmer}
\end{equation}
and is the difference between the load power and the total power production availability (given by the sum of the power produced by the PV plant and the available power of the \ac{bess}, $P_{\rm dis}$). If the control variable is negative, at least one diesel generator has to be switched on.

\begin{figure}[ht]
\centering
\includegraphics[width=0.8\columnwidth]{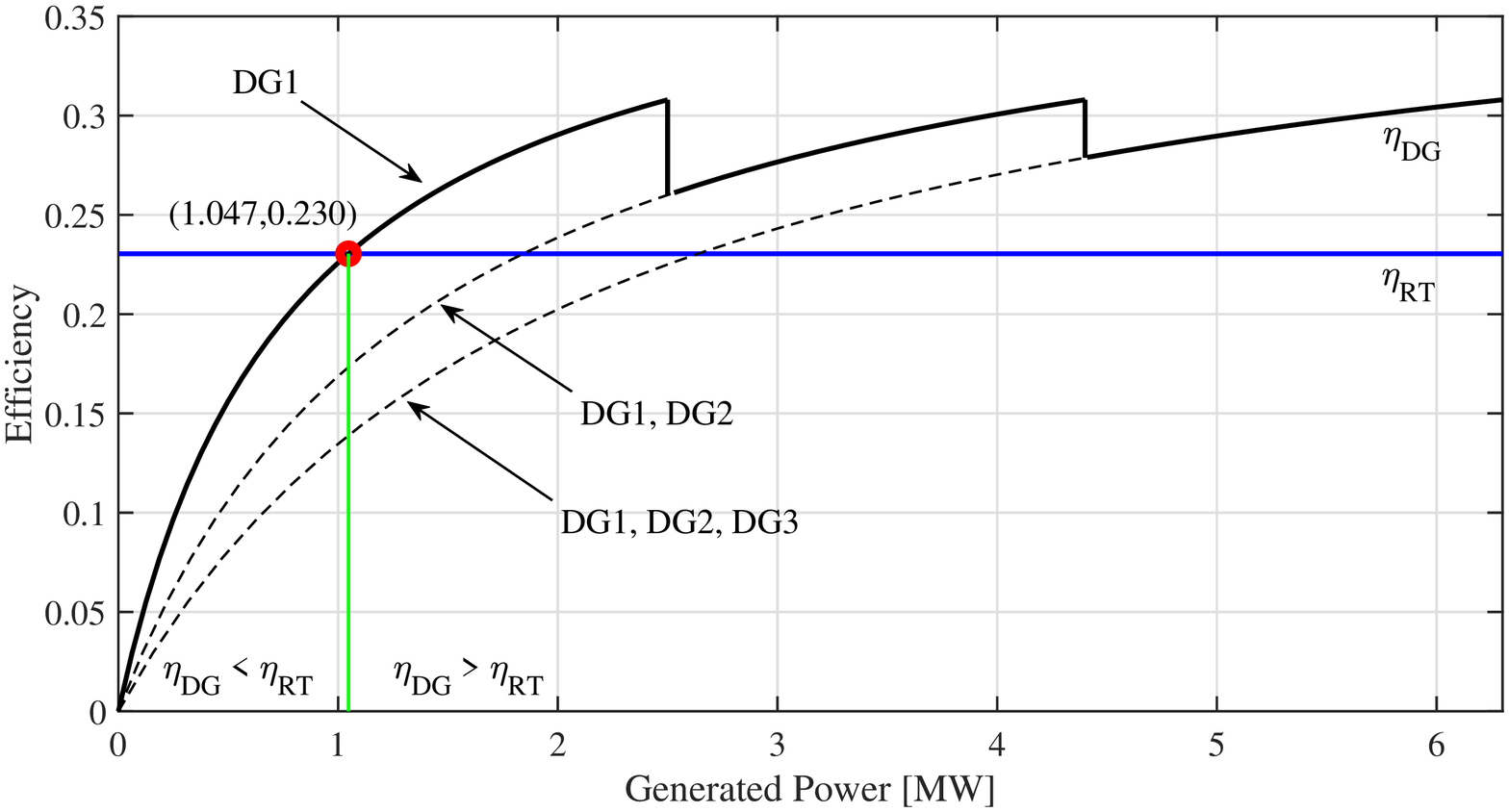}
\caption{Global efficiency curve of \acp{dg} with optimal dispatching.}
\label{fig:eff}
\end{figure}

The \ac{pms} has been designed to maximize the efficiency of power production. For this reason a further aspect has been analyzed and included in the algorithm. Figure~\ref{fig:eff} shows the global efficiency curve of the diesel power station ($\eta_{\rm DG}$), assuming to operate generators with the same incremental costs, which is the optimal solution of the multiple-machine dispatching problem. The blue line shows the round-trip efficiency curve ($\eta_{\rm RT}$) of the energy produced by the \acp{dg}, and provided by the \ac{bess} through a double conversion stage (charging and discharging). The round-trip efficiency is
\begin{equation}
    \eta_{\rm RT}= \bar{\eta}_{\rm DG} \eta_{\rm ch} \eta_{\rm dis}
    \label{eq:rt}
\end{equation}
where:
\begin{itemize}
    \item $\bar{\eta}_{\rm DG}$ is the maximum DG efficiency;
    \item $\eta_{\rm ch}$ is the global charging efficiency of \ac{bess};
    \item $\eta_{\rm dis}$ is the global discharging efficiency of \ac{bess}.
\end{itemize}

If the power requested to the \acs{dg} is lower than the point of intersection between the two efficiency curves (at \SI{1.047}{\mega\watt}), then it is convenient to utilize the diesel generator at the nominal power, i.e., the \ac{mcr}, and to store the energy excess in the battery.

Table~\ref{tab:states} illustrates the operational states of the \ac{pms}, i.e., the operating condition of the microgrid.

\begin{table}[h]
\centering
  \caption{Operation states of the microgrid: $P_{{\rm DG}_i}^{\rm n}$ are the nominal active powers of the \acp{dg}; $P_{{\rm ctrl}_i}$ are the dispatched active powers; $P_{{\rm ctrl}_i}^*$ denotes that the nominal power of the corresponding \ac{dg} is increased by 30\%.}
  \label{tab:states}
  \includegraphics[width=\columnwidth]{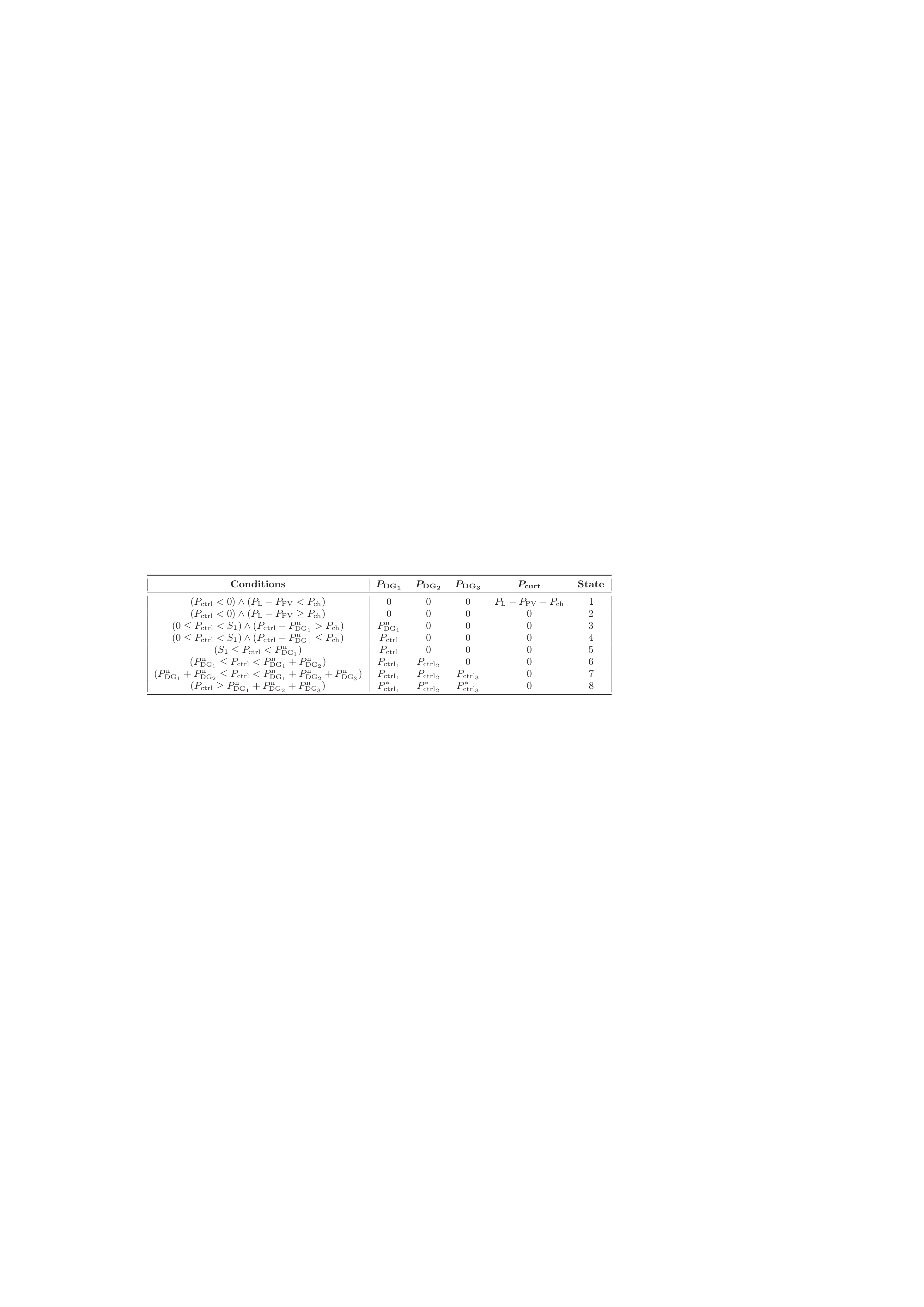}
\end{table}

\section{Long-Term Simulations}\label{sec:LTS}
In order to validate the proposed power management strategy quasi-dynamic simulations have been carried out. These simulations explore a one-year time span of operation of the microgrid depicted in Figure~\ref{fig:topology} through subsequent power flows in which the \ac{soc} is used as a ``state variable'' throughout the whole simulation. The parameters are reported in Table~\ref{tab:params}.

\begin{table}[h]
\caption{PMS study case parameters.}
\centering
\begin{tabular}{ll}
\toprule
\multicolumn{1}{c}{\textbf{Parameter}}  & \multicolumn{1}{c}{\textbf{Value}} \\
\midrule
$T_{\rm ctrl}$                  & \SI{5}{\minute}      \\
${\rm SOC}$ margin  ($m$)             & 0.05 (5\%)     \\
$\cos{\varphi}_{\rm DG}$               & 0.9            \\
$\cos{\varphi}_{\rm PV}$               & 1              \\
Expected grid losses ($k_{\rm loss}$)       & 2.6\%          \\
Voltage set-point ($V_{\rm BESS}$) & 1.025 p.u.     \\
\bottomrule
\end{tabular}
\label{tab:params}
\end{table}

The efficiency of the \ac{bess}, $\eta_{\rm BESS}$, i.e., the global charging/discharging efficiency of the storage system, including auxiliary loads, inverter, battery and filter losses is
\begin{equation}\label{eq6}
    \eta_{\rm BESS} =
    \begin{cases}
        \eta_{\rm ch}=0.87 & \text{if  $P\leq0$ \ \ (charging)}\\
        \frac{1}{\eta_{\rm dis}}=\frac{1}{0.86} & \text{if  $P>0$ \ \ (discharging)}.
    \end{cases}
\end{equation}
The values in \eqref{eq6} are extracted from a real measurement campaign~\cite{Conte_BESS}, and have been adopted for the current study case.

\begin{figure}[ht]
\centering
\includegraphics[width=0.9\columnwidth]{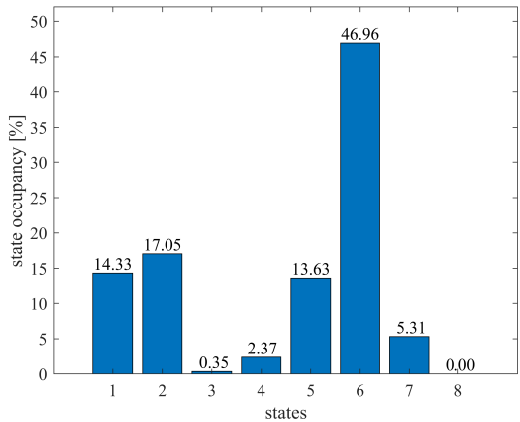}
\caption{State occupancy of the control during the one-year simulation.}
\label{fig:permanenza}
\end{figure}

In Figure~\ref{fig:permanenza} the occurrence of each state (as defined in Table~\ref{tab:states}) is reported. It describes the amount of time the microgrid spend in each of the eight states. It can be clearly noticed that the overload state (no. 8) is never visited. Moreover, the most probable state is no. 6, in which only two \acp{dg} are active.

\section{Conclusion}\label{sec:Conclusion}
This work presented the optimal sizing of an islanded microgrid and its management strategy. This strategy is based on an efficiency logic aimed to optimize the energy stored in the \ac{bess} and the energy produced by the renewable energy sources by minimizing the intervention of the fossil-fuel-based generators. The management strategy has been tested (and its effectiveness has been validated) by means of a one-year quasi-dynamic simulation.
Further work will be carried out to prove the PMS capability under different load profiles and component sizing.
\vspace{17.5pt}

\end{document}